\documentclass[aps,reprint,groupedaddress,superscriptaddress]{revtex4-1}
\usepackage{amsmath}
\usepackage{amssymb}
\usepackage{amsfonts}
\usepackage[dvipdfmx]{graphicx,color}
\usepackage{textgreek}
\usepackage{makecell,tabularx}
\setcellgapes{3pt}

\usepackage{amsfonts}
\usepackage[colorinlistoftodos]{todonotes}
\usepackage{multirow}

\usepackage[normalem]{ulem}
\usepackage{color}

\begin{document}

\title{Designing Leidenfrost drops}

\author{F. Pacheco-V\'azquez}
\affiliation{Instituto de F\'isica, Benem\'erita Universidad Aut\'onoma de Puebla, A. P. J-48, Puebla 72570, Mexico \\
}

\author{M. Aguilar-Gonz\'alez}
\affiliation{Instituto de F\'isica, Benem\'erita Universidad Aut\'onoma de Puebla, A. P. J-48, Puebla 72570, Mexico \\
}

\author{L. Victoria-Garc\'ia}
\affiliation{Instituto de F\'isica, Benem\'erita Universidad Aut\'onoma de Puebla, A. P. J-48, Puebla 72570, Mexico \\
}

\date{\today}

\begin{abstract}
Large Leidenfrost drops exhibit erratic bubble bursts to release vapour accumulated beneath the liquid,  becoming amorphous and unstable.  Here we report an original and remarkably simple method to stabilize and design a Leidenfrost puddle.  When a thin hydrophilic layer with a suitable design is placed over the liquid, the puddle adopts the layer shape due to adhesive forces and becomes stable.  We show a wide variety of designs obtained using this method.  Moreover, we determine the required layer dimensions to stabilize puddles,  the wetting and floating conditions, and a model for the evaporation rate of the designed drops.  Finally,  potential applications are highlighted.
\end{abstract}


\maketitle
Leidenfrost drops are extremely mobile because the vapour cushion suppresses contact with the substrate and friction can be neglected \cite{Leidenfrost1756}.  Thus,  a slight surface inclination or any small instability can power the droplet motion.  For instance,  small droplets \cite{Graeber2021} (or hydrogel spheres \cite{Waitukaitis2017}) deposited gently on rigid surfaces experience self-induced spontaneous oscillations and start bouncing with increasing heights.  The droplet motion can also be induced by using chemically heterogeneous surfaces \cite{Li2023} or placing the liquid on a hot ratchet \cite{Linke2006},  but even on a polished horizontal surface, a small droplet initially at rest self-propels without external field due to asymmetries in the vapour profile beneath the drop \cite{Bouillant2018}.  The stability and shape of a Leidenfrost drop depends on its capillary length: $\lambda_c = \sqrt{\sigma/\rho_l g}$,  where $\sigma$ and $\rho_l$ are surface tension and density of the liquid, respectively,  and $g$ is the acceleration of gravity \cite{Biance2003}.  For instance,  for water at boiling temperature, $\lambda_c \sim 2.5$ mm; droplets with radius smaller than this value adopt a quasi-spherical shape.  Larger drops are flattened by the effect of gravity,  which also induces the formation of a vapour blister beneath the liquid that eventually arises forming a central chimney that destabilizes the drop.  For very large puddles, several vapour chimneys can appear simultaneously,  leading to amorphous and unstable profiles \cite{Pomeau2012,Sobac2015,Hidalgo2016}. 

Recent research has been focused on potential applications of Leidenfrost drops in cooling processes \cite{Chabicovsky2015,Jiang2022}, drop transport \cite{Luo2017,Dodd2019} and selective coalescence \cite{Pacheco2021}.  For that reason,  there is a growing interest in preventing instabilities in Leidenfrost state and controlling the drop motion \cite{Vakarelski2012,Cousins2012,Jiang2022,Manjarik2023,Li2023,Yang2023}.  Nevertheless,  most techniques imply the modification of the substrate by physical and/or chemical treatment that can be expensive and difficult to achieve, or techniques applied only to microliter droplets \cite{Li2023,Yang2023}.  We discovered that it is possible not only to stabilize but also design the shape of a Leidenfrost puddle by gently placing a solid thin layer on the liquid.  If the layer is hydrophilic,  the liquid wets its surface and adopts its shape,  becoming also remarkably stable.  Although thin layers of different materials can be used,  we chose aluminum sheets that resist humidity,  temperature and are also easy to manipulate.  The method can be used to optimize the puddle evaporation rate, and we show its successful applicability in a cooling process.
\begin{figure}[hbt!]%
\centering
\includegraphics[width=0.47\textwidth]{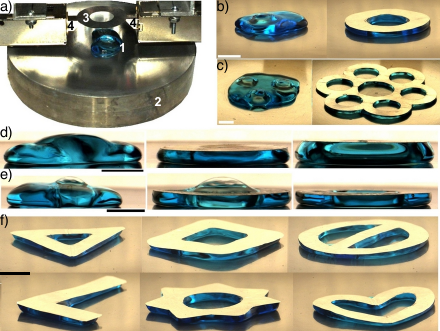}
\caption{a)  Experimental setup: a water puddle$^1$ levitates on a hotplate$^2$ when an aluminum layer of a given shape$^3$ is released on the droplet using synchronized electric gates$^4$.    b-c)  Snapshots showing how puddles of b) 1.5 ml and c) 5 ml with one and various vapour chimneys are shaped and stabilized with annular layers.  d-e)  Lateral views of puddles covered with d) a disk of diameter $D= 30$ mm,  and e) an annulus of external and internal radius $R_e=18$ mm and $R_i=9.8$ mm.  f) Puddles stabilized using layers with different shapes.  The scale bar represents 10 mm in all cases.}\label{fig1}
\end{figure}

\begin{figure*}%
\centering
\includegraphics[width=\textwidth]{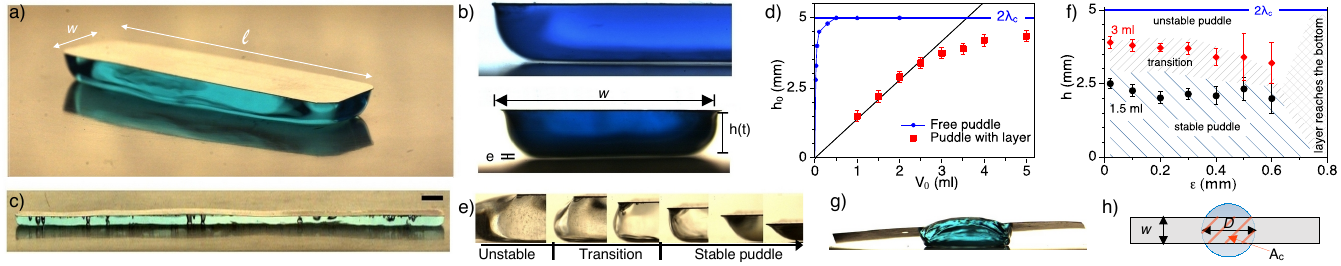}
\caption{a) Rectangular layer ($\textit{l} \times \textit{w} = 60 \times 12$ mm$^2$) above a 1.5 ml puddle that adopts the layer shape.  b) Lateral views of the puddle showing that the liquid perfectly wets the thin metal layer.  c) Long layer of $\textit{l} = 200$ mm and $ \textit{w}= 10$ mm stabilizing  a 6 ml puddle (scale bar = 10 mm).  d) Comparison of the initial height $h_0$ of the puddle with the $60 \times 12$ mm$^2$ layer (red squares) and without the layer (blue dots) depending on the liquid volume.  The black line corresponds to $h_0 = V_0/lw$.  e) Snapshot showing the evolution of a 3.5 ml puddle covered with a $60 \times 12$ mm$^2$ layer from an unstable to a stable phase.  f) Phase diagram of the puddle stability depending on the amount of liquid (represented by $h$) and the layer thickness $\varepsilon$.  g) Lateral image of the layer falling on the puddle.  h)  Top-view sketch showing the contact area $A_c \sim D w$ of the layer with the puddle.}\label{fig2}
\end{figure*}

\textit{Methodology.-} The procedure to design Leidenfrost drops is very simple. First, we deposit a desired volume of deionized water (1-10 ml) on a polished aluminum plate at temperature $T$ above the Leidenfrost temperature $T_L \sim 200^{\circ}$C.  The liquid forms an unstable Leidenfrost puddle characterized by one or more vapour chimneys appearing randomly across its surface.  Then,  a thin aluminum layer of thickness $\varepsilon$ ranging from 20 to 800 $\mu$m is released just above the puddle using two synchronized gates (see setup in Fig. \ref{fig1}a). The layer falls on the liquid and the puddle rapidly adopts the desired shape due to adhesion of the liquid to the sheet.  The process is filmed with a high speed camera at 125 fps to capture the wetting dynamics and subsequent evaporation. The simplicity of the method is better visualized in Movie 1 of the Supplementary material \cite{SI},  where a detailed description of the experimental setup is also provided. 

\textit{Requirements for design.-} Figures \ref{fig1}b-c show an annular layer shaping a 1.5 ml puddle and a multi-annular design sculpting a 5 ml puddle,  respectively. 
The use of layers with inner open areas allows to release the vapour more efficiently.  Figure \ref{fig1}d shows lateral view snapshots for the case of a disk without the central hole.  Even when the puddle adopts the layer shape,  the accumulation of vapour below the layer forms a bubble that cannot burst, generating a pulsating puddle of variable height.  In contrast,  when we place an annulus of the same area (Fig. \ref{fig1}e),  a bubble of vapour appears but it easily bursts and generates a steady chimney that releases the vapour continuously,  forming thereafter a stable donnut-shaped puddle.   Figure \ref{fig1}f shows that our method can be used to design puddles with a diversity of geometries. Nevertheless, to obtain stable designs some size requirements must be taken into account.  For instance,  the minimum inner diameter of the annulus in Fig. \ref{fig1}e must be approximately $15$ mm to allow chimney bursting, because this is roughly the mean diameter of erratic bubbles formed in a free puddle at the same temperature (see Supplementary Material \cite{SI}).  On the other hand,  in Ref. \cite{Biance2003},  it was found that bubble instabilities appear in water puddles of radius larger than $4 \lambda_c \sim 10$ mm.  Thus,  the layer design must have at least one dimension smaller than this value to release the vapour and avoid the accumulation of the gas beneath the puddle.  All the designs shown in Fig.  \ref{fig1}f satisfy this condition.  To support this point,  we used rectangular layers with different widths, $w$,  and lengths, $l$, in the range $w=5-40$ mm and $l=40-200$ mm.  In this geometry, the minimum path for vapour expulsion is $w/2$.  Figure \ref{fig2}a shows the case for $\textit{l} = 60$ mm and $\textit{w}=12$ mm. Note from this image and the lateral pictures in Fig. \ref{fig2}b that the liquid puddle is stable and wets all the covering layer from below,  adopting the rectangular shape.  In principle,  there is no upper limit for the length of the rectangular layer if $w/2 < 4 \lambda_c$.  A long layer of $l= 200$ mm and $w= 10$ mm shown in Fig. \ref{fig2}c remains incredibly stable covering a 6 ml water puddle that adopts the elongated ribbon shape.

\begin{figure*}
\centering
\includegraphics[width=0.95\textwidth]{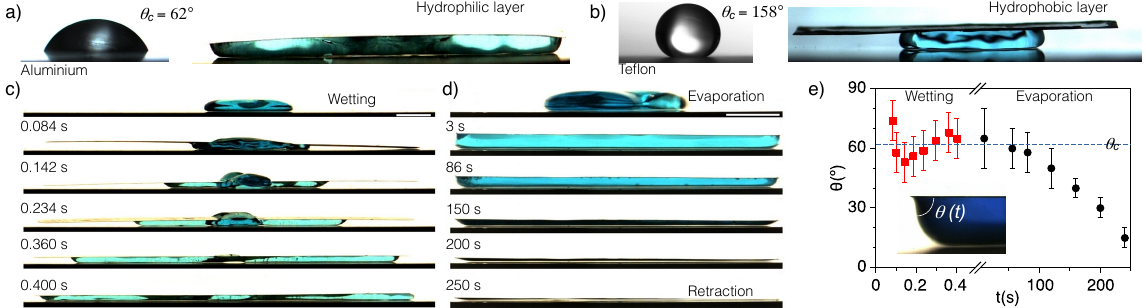}
\caption{ a) Contact angle $\theta_c=62\pm 4^{\circ}$ of a 10 $\mu$l water droplet deposited on the aluminum layer. Using this hydrophilic material,  the puddle adopts the layer shape.  b) When the layer is coated with Teflon, $\theta_c=158\pm 6^{\circ}$. This highly hydrophobic surface is not wetted by the puddle.  c-d) Snapshots showing  the fast wetting process of the hydrophilic rectangular layer on a 2.5 ml water puddle (c),  and the subsequent evaporation process (d) for a layer of $l=60$ mm and $w=12$ mm.  e) Evolution of the wetting angle $\theta$  with the upper aluminum layer,  measured as shown in the inset during wetting and evaporation phases.}
\label{fig3}
\end{figure*}

The size of the layer must also depend on the amount of water to be covered.  To take this aspect into account,  let us recall that the maximum height $h_0$ of a free puddle is $2\lambda_c \approx 5$ mm \cite{Biance2003,Sobac2015}.  In Fig. \ref{fig2}d,  we show that a free puddle reaches $h_0=2\lambda_c$ for initial volumes $V_0 > 0.5$ ml (blue dots).   When the puddle is covered by the aluminum layer of $\textit{l} = 60$ mm and $\textit{w}=12$ mm,  its height $h_0$ is reduced to a smaller value depending on the volume of liquid (red squares).  For $V_0< 3$ ml,  the puddle height can be estimated assuming a parallelepiped volume with $h_0 = V_0/\textit{lw}$ (black line),  indicating that all the liquid remains practically bellow the layer, as the one shown in Fig. \ref{fig2}a.  Notice in Fig. \ref{fig2}d that this is clearer when $h_0$ is similar or smaller than $\lambda_c=2.5$ mm.  At larger volumes,  $h_0$ starts to saturate to the maximum value $2\lambda_c$,  the liquid is spread horizontally surpassing the layer area and becomes unstable due to vapour accumulation beneath the layer. This is shown in  Fig. \ref{fig2}e for a large unstable puddle of $V_0=3.5$ ml.  As the liquid evaporates, the puddle contracts and a transition to a stable design occurs when the liquid is covered by the layer.  After that, the puddle remains stable until its total evaporation.  Figure \ref{fig2}f indicates these stability phases as the puddle height $h$ decreases due to evaporation, depending on the layer thickness $\varepsilon$.  A puddle of $V_0>3$ ml (red diamonds) is unstable until its height becomes similar to $\lambda_c$.  For $V_0=1.5$ ml (black dots) the puddle is stable since the beginning because $h_0 < \lambda_c$ and surface tension dominates against instabilities induced by gravity.

Note in Fig. \ref{fig2}f that  the layer sinks into the puddle if $\varepsilon \gtrsim 0.7$ mm, a value considerably smaller than the one expected for the aluminum layer to sink on a water pool ($\gtrsim 2.6$ mm,  see Supplementary material \cite{SI}).  However,  in our case,  the layer of area $lw$ falls on a finite free puddle,  see Fig. \ref{fig2}g,  and the initial contact area $A_c \sim Dw$ is roughly a fraction of the cross section of a puddle of diameter $D \sim (2 V_0 / \pi \lambda_c)^{1/2}$ and height $2\lambda_c$, as sketched in Fig. \ref{fig2}h.  The layer will float if its weight $mg$ is balanced by the buoyant force $F_b$ plus the capillary force $F_c$ \cite{Keller1998}.  Since $F_c$  is maximal when it has only a vertical component along the triple contact line \cite{Vella2015, Bormashenko2016},  the equilibrium condition can be written as: $\rho_a lw \varepsilon_{max} g  = \rho_l Dw (\sqrt{2}\lambda_c + \varepsilon_{max}) g + 2D \sigma $,  where $ \sqrt{2}\lambda_c$ is the depth of the contact line before sinking \cite{Naylor2022,Guo2018}.  Solving for  $\varepsilon_{max}$ one obtains: $\varepsilon_{max}=  2\lambda_c (\lambda_c/w + \sqrt{2}/2 )/\left[(\rho_a/\rho_l)(l/D) -1\right].$ In our experiments $l=60$ mm, $w=12$ mm,  $\lambda_c=2.5$ mm,  $\rho_a=2700$ kg/m$^3$ and $\rho_l =958$ kg/m$^3$ at $T_B=93^{\circ}$C (Puebla, Mexico,  2200 m.a.s.l \cite{Cervantes2020}).  For a puddle of 2 ml,  one obtains $\varepsilon_{max} \sim 0.7$ mm, in agreement with the limiting values found in Fig. \ref{fig2}f.

\textit{Wetting conditions.-} Another important aspect to consider is the wettability of the covering layer.  Figure \ref{fig3}a shows a sessile droplet of 10 $\mu$l deposited on the aluminum surface.  The contact angle $\theta_c\approx 62^{\circ}$ (determined using low bond axisymmetric drop shape analysis \cite{Stalder2010})  reveals a hydrophilic material that is totally wetted when the layer is deposited on the puddle.  In contrast,  when the same layer is coated with Teflon and the measurements are repeated,  $\theta_c \approx 158^{\circ}$, which indicates a highly hydrophobic surface, see Fig. \ref{fig3}b.  In this case,  the puddle keeps its unstable form below the covering layer.  Therefore,  it is necessary the use of hydrophilic layers to design Leidenfrost puddles (Movie 2 \cite{SI}).  Figure \ref{fig3}c shows in detail the wetting process for the hydrophilic case.  Following the contact, the liquid spreads horizontally toward the ends of the layer adopting the elongated shape in less than one second.  After spreading,  we show the evaporation process in Fig. \ref{fig3}d with a considerably longer time scale $\tau \sim 250$ s.  During this time,  the puddle height $h(t)$ decreases continuously until forming a very thin liquid film that retracts previous to the total evaporation.  During the wetting phase,  the angle $\theta$ between the layer and the liquid-air interface (see Fig. \ref{fig3}e) is an advancing angle that takes values fluctuating around $\theta_c$.  During the evaporation phase, $\theta<\theta_c$ is a receding angle that decreases with time due to the reduction of the liquid volume anchored to the hydrophilic layer by adhesive forces.

\begin{figure}[h]
\centering
\includegraphics[width=0.5\textwidth]{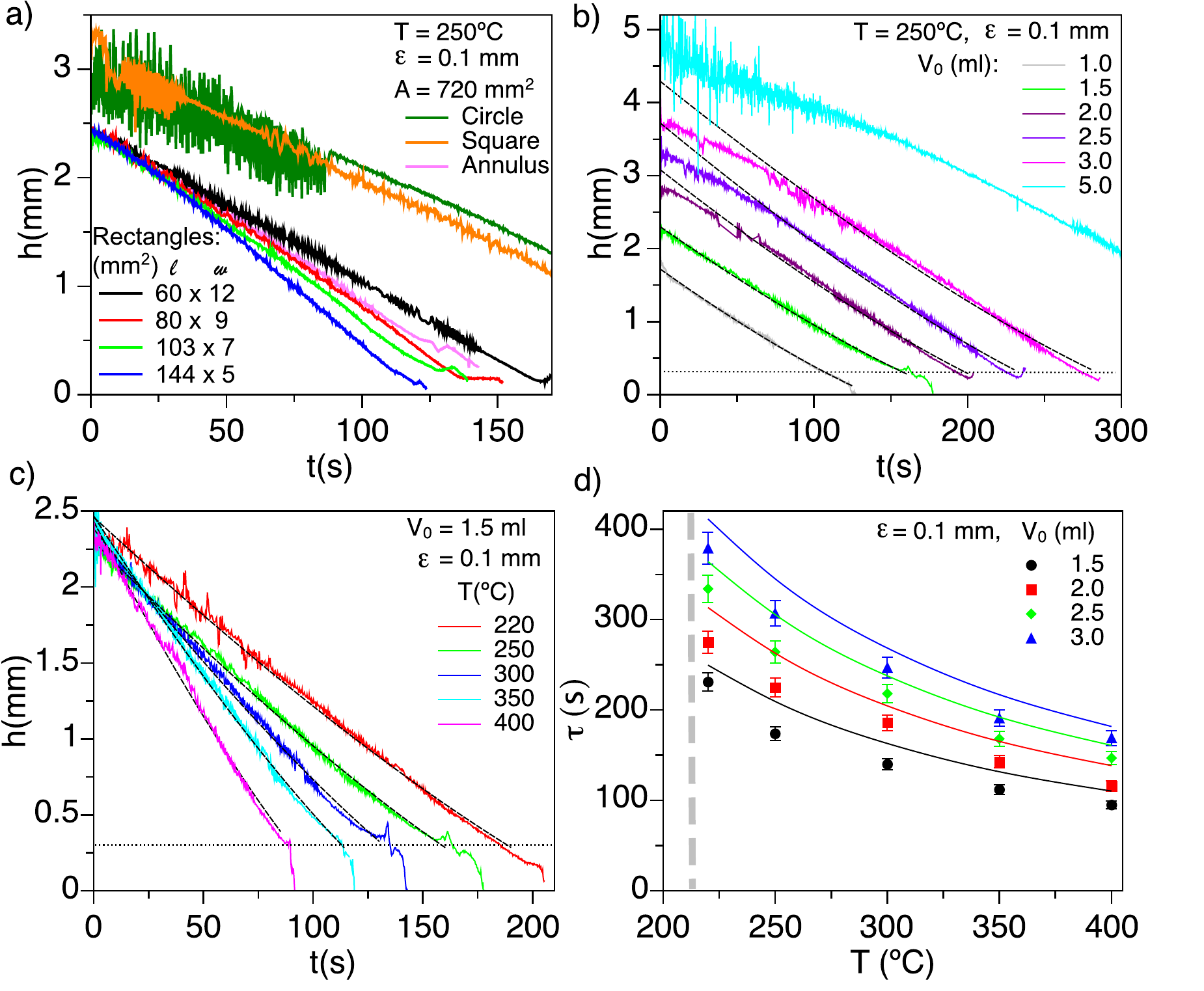}
\caption{Puddle heigh $h$ as a function of time $t$ (colour lines) depending on a) the geometry, b) the initial volume $V_0$,  and c) the plate temperature $T$ (dashed lines correspond to the model discussed in the text).  d) Total evaporation time $\tau$ of different liquid volumes (solid points) on plates at different temperatures $T >T_L$ (colour lines correspond to the model prediction,  and the  vertical dashed line indicates $T=T_L$).}\label{fig4}
\end{figure}

\textit{Evaporation process.-}  Let us now determine how the evaporation process depends on the geometry of the layer,  initial liquid volume $V_0$ and plate temperature $T$.  Figure \ref{fig4}a shows $h(t)$ for puddles of $V_0=1.5$ ml covered with aluminum sheets of different shapes but with the same mass ($m=0.21$ g),  thickness ($\varepsilon=0.1$ mm) and area ($A \approx 720$ mm$^2$).  For annular and rectangular geometries that allow central or lateral vapour expulsion,  $h(t)$ starts from $h_0 \approx 2.4$ mm $\sim V_0/lw$; however,  for circular and square geometries $h_0 \sim 3$ mm, indicating that big vapour bubbles are trapped beneath the metal sheet,  as in Fig. \ref{fig1}d.  These bubbles are responsible of the large fluctuations of $h(t)$ observed in Fig. \ref{fig4}a for those geometries.  For rectangular layers,  the evaporation rate $dh/dt$ increases (more pronounced slopes) when $w$ is decreased,  in agreement with the hypothesis that vapour can escape more easily following the shorter lateral path.  Therefore,  long and narrow geometries optimize the evaporation process.  Figure \ref{fig4}b shows $h$ vs $t$ for puddles of different volumes covered with the same rectangular layer ($\varepsilon=0.1$ mm, $l=60$ mm, $ w=12$ mm) and Fig.  \ref{fig4}c shows the effect of the plate temperature.   As one expects, $dh/dt$ increases with $T$,  leading to a decrease of the evaporation time $\tau$,  as it is summarized in Fig. \ref{fig4}d for different volumes.

To model the results shown in Figs. 4b,c,d,  let us consider that the evaporation rate due to the heat transfer from the hot plate to the liquid through the vapour layer of thickness $e(t)$  is given by $dm_v/dt= \kappa \Delta T A / e L$ \cite{Gottfried1966,Biance2003},
where $\kappa$ is the thermal conductivity of the vapour,  $L$ the latent heat of evaporation, and $\Delta T= T - T_B$.  On the other hand,  due to the hydrostatic pressure of the levitating liquid $P_l= \rho g h(t)$ plus the pressure of the metal sheet $P_a =  \rho_a g \varepsilon$,  the generated vapour (with viscosity $\mu_v$ and density $\rho_v$) flows between the hot plate and the puddle travelling the shorter distance $\Delta x \approx w/2$,  to be expelled laterally from both sides of length $l$. This situation can be approached assuming a planar Poiseuille flow imposed by the pressure gradient $\Delta P/\Delta x$ \cite{Kundu},  where $\Delta P =  P_l + P_a$.  Under these assumptions,  the mass flow rate of vapour is given by $dm_v/dt = \rho_v l  \Delta P e^3/ 3 \mu_v w$.
Equalising both expressions for $dm_v/dt$,  an expression for the vapour film thickness can be found  \cite{SI}. From mass conservation,  $dm_v/dt = -\rho_l lw (dh/dt)$,  leading to $dh/dt = -\rho_v \Delta P e^3/ 3\mu_v \rho_l w^2$.  Integrating one obtains:
\begin{equation}
  h(t) = \frac{1}{\rho_l g} \left[ C_T t + C_h \right]^{4/3} - \frac{\rho_a }{\rho_l}\varepsilon,  
  \label{Eq1}
\end{equation}
where $C_T = - (\rho_v g/4\mu_v w^2)(3\mu_v \kappa\Delta T w^2 / \rho_v L)^{3/4}$  and $C_h = \left(\rho_l gh_0 +  \rho_a g \varepsilon \right)^{3/4}$.  We used  Eq. \eqref{Eq1} to fit the experimental data shown in Figs. \ref{fig4}b-c (dashed black lines).  It can be noticed that the fitting is better for small stable volumes,  where the laminar Poiseuille flow is valid.  The values of  $\kappa$,  $\mu_v$ and $\rho_V$ depending on temperature \cite{Engtoolbox2004,Engtoolbox2018} can be used to calculate $C_T$ and $C_h$ (see supplementary material \cite{SI}).  Then,   $\tau(T,V_0)$ can be obtained from Eq.  \eqref{Eq1} considering $h(\tau) =0$, which leads to: $\tau = \left[(\rho_a g \varepsilon)^{3/4} - C_h)\right]/C_T$.  The computed values of $\tau(T,V_0)$ represented by solid lines in Fig. \ref{fig4}d follow the same trend than the experimental data, and it that sense the model is successful. However,  the experimental values are overestimated because the model assumes complete wetting during the whole process, while in the experiments there is a retraction phase characterized by the sudden decrease of $h(t)$ below the horizontal dashed lines in Figs \ref{fig4}b-c.  Moreover,  we neglect the evaporation from the lateral face of the puddle,  which accounts for $10\%$ of the total evaporation in Leidenfrost drops \cite{Sobac2015}.  

\begin{figure}[h!]
\centering
\includegraphics[width=0.48\textwidth]{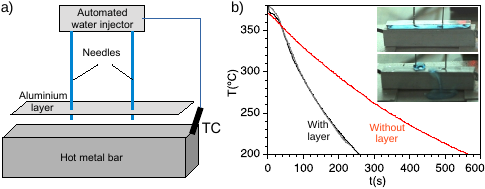}
\caption{a) Design of a Leidenfrost cooler adapted to a metal bar.   b) Change of the bar temperature with time with and without the aluminum layer in a real experiment (see text). }\label{fig5}
\end{figure}

Finally,  the fact that the puddle attaches to the layer and adopts its shape can be exploited for potential applications.  In Fig. \ref{fig5}, we depict  a cooling device adapted to a hot rectangular bar of mass $m_b=32$ g and heat capacity $C_b=0.217$ Cal/g $^\circ$C,  which is cooled down from $T_0=370^{\circ}$C to $T_F=200^{\circ}$C by placing a thin aluminum layer on the bar and adding 2.5 ml of water through needles with a water injector (the required water volume can be estimated by $V_0=m_b C_b (T_0-T_F)/\rho_l  L$).  As shown in Fig. \ref{fig5}b,  the temperature is reduced considerably faster respect to the process without the  layer.  The upper snapshot shows the stable cooling mechanism using the layer, see also Movie 1 \cite{SI};  without the layer (bottom image),  the liquid undesirably spills from the bar and is wasted.

In summary,  we propose a novel  and surprisingly simple technique to stabilize and design Leidenfrost puddles based on the adhesion of liquid to a hydrophilic thin sheet deposited on the puddle.  The geometry of the layer must be smartly designed to avoid vapour accumulation,  and we provide the required conditions to achieve this goal.  The versatility of the method allows to imagine a broad variety of potential applications,  including efficient cooling processes,  controlled propulsion (using magnetic layers) and low-friction machines.

\paragraph*{Acknowledgements.-}
This research was supported by CONAHCYT-Mexico and VIEP-BUAP Project 2024.

\end{document}